\documentclass[preprint,pra,nofootinbib]{revtex4}
\usepackage{mathrsfs}
\usepackage{graphicx}
\usepackage{epsfig}
\usepackage{dcolumn}% Align table columns on decimal point
\usepackage{bm}% bold math
\usepackage{amsmath,amssymb,amsthm}
\usepackage[colorlinks=true,linkcolor=blue]{hyperref}
\usepackage{subfigure}
\usepackage{booktabs}
\usepackage[mathscr]{euscript}
\usepackage{ulem}

\newcommand{\bea}{\begin{eqnarray}}
\newcommand{\eea}{\end{eqnarray}}
\newcommand{\beq}{\begin{equation}}
\newcommand{\eeq}{\end{equation}}

\def\/{\over}

\begin{document}
\title{Exact parity and time reversal symmetry invariant and symmetry breaking solutions for a nonlocal KP system}
\author{Wenbiao Wu, S. Y. Lou (Corresponding author)\footnote{Corresponding author: S. Y. Lou; email: lousenyue@nbu.edu.cn; Institute: Center for Nonlinear Science and Department of Physics, Ningbo University, Ningbo, 315211, China}}
\affiliation{\footnotesize{Center for Nonlinear Science and Department of Physics, Ningbo University, Ningbo, 315211, China} }

\begin{abstract}
 A nonlocal Alice-Bob Kadomtsev-Petviashivili (ABKP) system with shifted-parities ($\hat{P}_s^x$ and $\hat{P}_s^y$  parities with shifts for the space variables $x$ and $y$) and delayed time reversal ($\hat{T}_d$, time reversal with a delay) symmetries is investigated. Some types of $\hat{P}_s^y\hat{P}_s^x\hat{T}_d$ invariant solutions including multiple soliton solutions, Painlev\'e reductions and soliton and $p$-wave interaction solutions are obtained via $\hat{P}_s^y\hat{P}_s^x\hat{T}_d$ symmetry and the solutions of the usual local KP equation. Some special $\hat{P}_s^y\hat{P}_s^x\hat{T}_d$ symmetry breaking multi-soliton solutions and cnoidal wave solutions are found from the $\hat{P}_s^y\hat{P}_s^x\hat{T}_d$ symmetry reduction of a coupled local KP system.
\end{abstract}

\maketitle

\section{Introduction}
In 2013, Ablowitz and Musslimani\cite{ref1} introduced a integrable nonlocal nonlinear Schr\"{o}dinger equation
\begin{equation}
iA_t+A_{xx}\pm A^2 B=0,\ B=\hat{f}A=\hat{P}\hat{C} A=A^*(-x,t),\label{AM}
\end{equation}
where the operators $\hat{P}$ and $\hat{C}$ are the usual parity and charge conjugation. Similarly, other types of nonlocal nonlinear PC, PT and PTC symmetric systems such as the coupled nonlocal NLS systems\cite{ref2}, the nonlocal KdV and modified KdV systems\cite{ref3,ref4,ref5}, the discrete nonlocal NLS systems\cite{ref6}, and the nonlocal Davey-Stewartson systems\cite{ref7,ref8,ref9} have been obtained because of the PTC symmetries. PTC symmetries are important in not only particle physics\cite{ref10}, but also many other physical fields such as optics\cite{ref11}, quantum field theory\cite{ref12}, electric circuits. Recently\cite{ref13}, PT-symmetric nonlocal NLS equation \eqref{AM} is used to describe the extension of properties of traditional macroscopic magnetic systems\cite{ref14}.

It is well known that there are various correlated and/or entangled events that may be happened in different times and places. To describe two-place physical problems, Alice-Bob (AB) systems \cite{ref15} are proposed by using the shifted parity ($\hat{P}_{s}$), delayed time reversal ($\hat{T}_{d}$) and charge conjugate ($\hat{C}$) symmetries. If one event (A, Alice event) is correlated/entangled another (B, Bob event), we denote the correlated relation as $B=\hat{f}A$ for suitable $\hat{f}$ operators. Usually, the event $A=A(x,\ t)$ happened at $\{x,\ t\}$ and event $B=B(x',\ t')$ happened at $\{x',\ t'\}=\hat{f}\{x,\ t\}$. In fact, $\{x',t'\}$ is usually far away from $\{x,\ t\}$. Hence, the intrinsic two-place models or Alice-Bob systems are nonlocal. In addition to the nonlocal nonlinear Schr\"odinger equation \eqref{AM} \cite{BYang}, there are many other types of two-place nonlocal models, such as the nonlocal KdV systems \cite{ref25}, the nonlocal modified KdV systems \cite{ref4,ref5}, the discrete nonlocal NLS systems \cite{ref6}, the coupled nonlocal NLS systems \cite{ref2} and the nonlocal Davey-Stewartson systems \cite{ref7,ref8,ref9}, etc.

In \cite{ref15}, one of us (Lou) proposed a series of integrable AB systems including the ABKdV systems, ABmKdV systems, ABKP systems, AB-sine Gordon (ABsG) systems, ABNLS systems, AB-Toda (ABT) systems and ABAKNS systems. Furthermore, by using the $\hat{P}^y_{s},\ \hat{P}^x_{s},\ \hat{T}_{d}$ and $\hat{C}$ symmetries, their $\hat{P}^y_{s},\ \hat{P}^x_{s},\ \hat{T}_{d}$ and $\hat{C}$ invariant multi-soliton solutions are obtained in elegant forms.

 In addition, Lou established a most general ABKdV equation and presented its $\hat{P}^y_{s},\ \hat{P}^x_{s},\ \hat{T}_{d}$ and $\hat{C}$ invariant Painlev\'e II reduction and soliton-cnoidal periodic wave interaction solutions for the ABKdV system \cite{ref16}. However, to find $\hat{P}^y_{s},\ \hat{P}^x_{s},\ \hat{T}_{d}$ and $\hat{C}$ symmetry breaking solutions is much more difficult.

In this paper, influenced by the idea of \cite{ref16,ref25} we investigate a special ABKP system with $\hat{P}^y_{s}\hat{P}^x_{s}\hat{T}_d$ symmetry. The $\hat{P}^y_{s}\hat{P}^x_{s}\hat{T}_d$ invariant solutions are obtained by applying the $\hat{P}^y_{s}\hat{P}^x_{s}\hat{T}_d$ symmetry to the solution of the local KP equation while the $\hat{P}^y_{s}\hat{P}^x_{s}\hat{T}_d$ symmetric breaking solutions are solved by introducing a coupled KP system.
\section{An ABKP system}
In this section, we study a special ABKP system
\begin{equation}
\begin{split}
&[A_t+A_{xxx}+\frac{3}{2}(A+B)(B_x+3A_x)]_x+\sigma^2A_{yy}=0,
\label{ABKP}\\
&B=\hat{P_s^x}\hat{P_s^y}\hat{ T_d}A=A(-x+x_0,\ -y+y_0,\ -t+t_0)
\end{split}
\end{equation}
with arbitrary constants $x_0,\ y_0$ and $t_0$.

The ABKP system Eq.(\ref{ABKP}) can be derived by applying the consistent correlated bang (CCB) approach to the usual KP equation
\begin{equation}
(u_t+12uu_x+u_{xxx})_x+\sigma^2u_{yy}=0\label{KP}
\end{equation}
as follows the reference \cite{ref16}. The KP equation \eqref{KP} firstly derived to study the evolution of long ion-acoustic waves of small amplitude propagating in plasmas under the effect of long transverse perturbations \cite{kp1}. The KP equation was widely accepted as a natural extension of the classical KdV equation to two spatial dimensions, and was later derived as a model for surface and internal water waves \cite{kp2}, and in nonlinear optics \cite{kp3} and almost in all other physical fields such as in shallow water waves, ion-acoustic waves in plasmas, ferromagnetics, Bose - Einstein condensation and string theory. The KP equation is also used as a classical model for developing and testing of new mathematical techniques.

The first step of CCB is to bang a single equation to a coupled system. The simplest way is substituting the ansatz $u=\frac{1}{2}(A+B)$ into the usual KP equation  \eqref{KP}
\begin{equation}
[A_t+B_t+A_{xxx}+B_{xxx}+6(A+B)(A_x+B_x)]_x
+\sigma^2(A_{yy}+B_{yy})=0
\end{equation}
which can be banged to two equations
\begin{eqnarray}
&&\left[A_t+A_{xxx}+\frac{3}{2}(A+B)(B_x+3A_x)\right]_x
+\sigma^2A_{yy}+G(A,B)=0,\nonumber\\
&&\left[B_t+B_{xxx}+\frac{3}{2}(A+B)(A_x+3B_x)\right]_x
+\sigma^2B_{yy}-G(A,B)=0,\label{Bang}
\end{eqnarray}
where $G(A,\ B)$ is an arbitrary functional of $A$ and $B$.

The second step of CCB is to take a correlation
\begin{equation}
B=\hat{f}A \label{B}
\end{equation}
between the fields $A$ and $B$ for a suitable operator $\hat{f}$ which can be taken as one of the elements of the eighth order $PTC$ group
\begin{equation}
{\cal{G}}=\{I,\ \hat{C},\ \hat{P}_s^x\hat{T}_d,\ \hat{P}_s^y,\  \hat{C}\hat{P}_s^x\hat{T}_d,\  \hat{C}\hat{P}_s^y,\  \hat{P}_s^x\hat{T}_d \hat{P}_s^y,\  \hat{C} \hat{P}_s^x\hat{T}_d\hat{P}_s^y\},\label{G}
\end{equation}
where the operators $\hat{C}$, $\hat{P}_s^x$, $\hat{T}_d$ and $\hat{P}_s^y$ are defined as
$$\hat{C}A=A^*,\ \hat{P}_s^xx=-x+x_0,\  \hat{P}_s^yy=-y+y_0,\  \hat{T}_dt=-t+t_0.$$
The last step of CCB is to fix the consistent condition under the correlation \eqref{B}.
Applying the correlation \eqref{B} on the banged system \eqref{Bang} will lead to a compatibility condition
\[G(A,B)=-\hat{f}G(A,B),\]
which means the arbitrary functional $G(A,B)$ should be $\hat{f}$ antisymmetric.

Generally there are numerous functionals satisfying $\hat{f}$ antisymmetric condition.
 For simplicity, in this paper, we discuss only for $G(A,B)=0$ and
$$\hat{f}=\hat{P}_s^x\hat{T}_d \hat{P}_s^y.$$
Thus, the banged system \eqref{Bang} is reduced to the special ABKP system \eqref{ABKP}.

\section{$\hat{P}_s^x\hat{P}_s^y\hat{T}_d$ invariant solutions to the ABKP systems}
As the ABKP system \eqref{ABKP} is directly derived from the usual KP equation \eqref{KP}, all the $\hat{P}_s^x\hat{P}_s^y\hat{T}_d$ invariant solutions of the KP equation \eqref{KP} are also solutions to the ABKP system \eqref{ABKP}. Various exact solutions of the KP equation \eqref{KP} have been studied in literature. Thus we can start from those known solutions of \eqref{KP} to select out the $\hat{P}_s^x\hat{P}_s^y\hat{T}_d$ invariant ones for the ABKP system \eqref{ABKP}. Here we list three special significant examples, the multiple solitons, the Painlev\'e reductions and the interaction solutions between soliton and $p$-waves including cnoidal periodic waves.
\subsection{$\hat{P}_s^x\hat{P}_s^y\hat{T}_d$ invariant multi-soliton solutions}
For the KP equation \eqref{KP}, it is well known that the multiple soliton solution possesses the form \cite{ref17}
\begin{equation}
u=(\ln F)_{xx}, \label{u}
\end{equation}
\begin{equation}
F=\sum_{\nu}\exp\left(\sum_{j=1}^N\nu_j\xi_j+\sum_{i\leq j\leq}^N\nu_j\nu_i\theta{ij}\right),\label{F}
\end{equation}
where the summation of $\nu$ should be done for all permutations of $\nu_i=0,\ 1,\ i=1,\ 2,\ \ldots,\ N$ and
\begin{equation}
\xi_j=k_jx+l_jy-k_j^{-1}\left(k_j^4
+\sigma^2l_j^2\right)t+\xi_{0j},\ \exp(\theta_{ij})=\frac{3k_i^2k_j^2(k_j-k_i)^2
-\sigma^2(l_jk_i-l_ik_j)^2}
{3k_i^2k_j^2(k_j+k_l)^2-\sigma^2(l_jk_i-l_ik_j)^2}. \label{xi}
\end{equation}
It is clear that the solution  \eqref{u} with  \eqref{F} and  \eqref{xi} is not $\hat{P}_s^x\hat{P}_s^y\hat{T}_d$ invariant. To find the N-soliton solution of the ABKP system  \eqref{ABKP}, we rewrite \eqref{xi} as
\begin{equation}
\xi_j=\eta_j-\frac{1}{2}\sum_{i=1}^{j-1}\theta_{ij}
-\frac12\sum_{i=j+1}^N\theta_{ij}
\end{equation}
where
\begin{equation}
\eta_j=k_j\left(x-\frac{1}{2}x_0\right)
+l_j\left(y-\frac{1}{2}y_0\right)-k_j^{-1}\left(k_j^4
+\sigma^2l_j^2\right)\left(t-\frac{1}{2}t_0\right)+\eta_{0j},
\label{eta11}
\end{equation}
then the N-soliton solution of the KP equation  \eqref{KP} becomes \cite{ref15}
\begin{equation}
u=\left[\ln\sum_{\nu=-1,1}
K_\nu\cosh\left(\frac12\sum_{j=1}^N\nu_j\eta_j\right)\right]_{xx},
\label{u12}
\end{equation}
where the summation of $\nu=\{\nu_1,\ \nu_2,\ \ldots,\ \nu_N\}$ should be done for all non-dual permutations of $\nu_i=1,\ -1,\ i=1,\ 2,\ \ldots,\ N$ ($\nu$ and $\nu'$ are dual if $\nu=-\nu'$), and

\[K_\nu=\prod_{i>j}\sqrt{3k_i^2k_j^2(k_i-\nu_i\nu_jk_j)^2
-\sigma^2(l_ik_j-k_il_j)^2}.\]

From the expression \eqref{u12}, it is easy to see that
\begin{equation}
A=u_{\eta_{0j}=0} \label{A13}
\end{equation}
solves the ABKP system \eqref{ABKP}.
The solution \eqref{A13} with \eqref{eta11} and  \eqref{u12} to the ABKP system \eqref{ABKP} is $\hat{P}_s^y\hat{P}_s^x\hat{T}_d$ invariant because $A=u_{\eta_{0j}=0}=\hat{P}_s^y\hat{P}_s^x\hat{T}_d u_{\eta_{0j}=0}$.
\subsection{$\hat{P}_s^y\hat{P}_s^x\hat{T}_d$ invariant Painlev\'e reduction solutions to the ABKP systems}
The knowledge of the symmetries is very useful to enhance our understanding of complex physical phenomena, to simplify and even completely solve the complicated problems. Furthermore, the study of symmetries has been manifested as one of the most important and powerful methods in almost every branch of science especially in physics and mathematics. It is particularly fundamental to find the symmetries of a nonlinear equation in the development of the theory of the integrable systems because of the existence of infinitely many symmetries\cite{ref18,ref28,ref29}.

 By using a direct method or nonclassical symmetry approach one can find that the usual KP equation \eqref{KP} possesses a symmetry reduction solution \cite{ref27}
 \begin{eqnarray}
 &&u=\frac1{12\theta^2}\left[-\theta\theta_tx
 -\theta\delta_t-\sigma^2\big(\theta^2_yx^2
 +2x\theta_y\delta_y+\delta_y^2\big)\right]
 +\frac12\theta^2w(z),\label{uw}\\
 && z=\theta x+\delta,\ \theta=\theta(y,\ z)\label{z}\\
 && \theta_{yy}=a\sigma^2\theta^5, \label{theta}\\
 &&\theta_t+\sigma^2\delta_{yy}=(a\delta+b)\theta^4\label{delta}
 \end{eqnarray}
 with two arbitrary constants $a$ and $b$ and the ordinary differential reduction equation
 \begin{equation}
w''''+6(w'^{2}+ww'')+(az+b)w'+2aw=\frac13(az+b)^2\label{w}
\end{equation}
which is equivalent to the Painlev\'{e} IV equation ($a\neq0$), Painlev\'e III equation ($a=0,\ b\neq 0$) and Painlev\'e II equation ($a=b=0$). All solutions of \eqref{theta} and \eqref{delta} have been listed in \cite{ref27}.

Therefore, in order to search $\hat{P}_s^y\hat{P}_s^x\hat{T}_d$ invariant reduction solutions for the ABKP system \eqref{ABKP},
Some special constraints have to be added for the solutions of \eqref{theta} and \eqref{delta}. For simplicity, we consider only the case of $x_0=y_0=t_0=0$ of \eqref{ABKP} for the Painlev\'e reduction solutions \eqref{uw}. From the equations \eqref{uw}, \eqref{z}, \eqref{theta}, \eqref{delta} and \eqref{w}, we can find the $\hat{P}^y\hat{P}^x\hat{T}$ invariant Painlev\'e reduction solutions
\begin{eqnarray}
 A=\frac1{12\theta^2}\left[-\theta\theta_tx
 -\theta\delta_t-\sigma^2\big(\theta^2_yx^2
 +2x\theta_y\delta_y+\delta_y^2\big)\right]
 +\frac12\theta^2w(z),\label{Aw}
 \end{eqnarray}
 for the ABKP system \eqref{ABKP} for $x_0=y_0=t_0=0$,
where $\theta$ and $\delta$ are solutions of \eqref{theta} and \eqref{delta} under the conditions
\begin{eqnarray}
 \theta=-\theta(-y,\ -t),\ \delta=\delta(-y,\ -t),\ b\neq0,\ \label{bneq0}
 \end{eqnarray}
or
\begin{eqnarray}
 \theta=\pm \theta(-y,\ -t),\ \delta=\mp \delta(-y,\ -t),\ b=0. \label{beq0}
 \end{eqnarray}

\subsection{Interaction solutions between soliton and $p$-wave of the ABKP systems}
Many authors have studied interaction solutions among different type of nonlinear excitations, especially, the soliton-cnoidal wave interaction solutions for many nonlinear systems such as the KdV equation and the KP equation \cite{ref30,ref26}. For example, one can directly prove that the KP equation \eqref{KP} possesses an interaction solutions between a soliton and a $p$-wave
\begin{eqnarray}
u&=&-f_{x}^2\tanh^2(f)+f_{xx}\tanh(f)
-\frac{1}{12}\frac{f_{y}^2\sigma^2}{f_{x}^2}
+\frac{2}{3}f_{x}^2\nonumber\\
&&-\frac{1}{3}\frac{f_{xxx}}{f_{x}}-\frac{1}{12}
\frac{f_{t}}{f_{x}}+\frac{1}{4}\frac{f_{xx}^2}{f_{x}^2},\ f=kx+ly+\omega t+p_0+p, \label{f}
\end{eqnarray}
where $k,\ l,\ \omega$ and $p_0$ are arbitrary constants and the $p$-wave satisfies
\begin{equation}
\left[\frac{\omega+p_t}{k+p_x}+\frac{2(k+p_x)p_{xxx}-3p_{xx}^2
+\sigma^2(l+p_y)^2}{2(k+p_x)^2}-2(k+p_x)^2\right]_x
+\sigma^2\left(\frac{l+p_y}{k+p_x}\right)_y=0. \label{p}
\end{equation}
For vanishing $p$-wave, $p=0$, the solution \eqref{f}
reduces back to the single soliton solution of the usual KP equation \eqref{KP}.

For traveling $p$-wave, the solution of \eqref{p} can be written as
\begin{equation}
p=\int W(X)\mbox{dX},\quad X=k_1x+l_1y+\omega_1t+X_0, \label{pW}
\end{equation}
with $W=W(X)$ being a solution of
\begin{eqnarray}
&&W_X^2=4W^4+c_3W^3+c_2W^2+c_1W+c_0, \label{Wx}\\
&&c_1=\frac{2kc_2}{k_1}-\frac{3c_3k^2}{k_1^2}
-\frac{2l_1\sigma^2}{k_1^5}(kl_1-k_1l)+\frac{16k^3}{k_1^3}
-\frac{k\omega_1}{k_1^4}+\frac{\omega}{k_1^3},\label{rc1}\\
&&c_0=\frac{k^2c_2}{k_1^2}-\frac{2c_3k^3}{k_1^3}
-\frac{\sigma^2}{3k_1^6}(kl_1-k_1l)(5kl_1+k_1l)
+\frac{12k^4}{k_1^4}
-\frac{k^2\omega_1}{k_1^5}+\frac{k\omega}{k_1^4},\label{rc0}
\end{eqnarray}
while $c_2,\ c_3,\ k,\ k_1,\ l,\ l_1,\ \omega,\ \omega_1$ and $X_0$ are all arbitrary constants.

Because of the M\"obious transformation invariance of \eqref{Wx}, its solution can be expressed by means of the Jacobi elliptic functions \cite{Phi3},
\begin{eqnarray}
&&W=\frac{a\ \mbox{\rm sn}(X,\ m)+b}{1+c\ \mbox{\rm sn}(X,\ m)},\ a=bc\pm \sqrt{(c^2-1)(c^2-m^2)}, \label{SW}
\end{eqnarray}
with arbitrary constants $b,\ c,\ m$ and the dispersion relation
\begin{eqnarray}
\omega_1=\frac{k_1}k\omega-\frac{k_1^5}{k^2}c_0+k_1^3c_2-2kk_1^2c_3
+12k_1k^2-\frac{\sigma^2}{3k_1k^2}(5kl_1+lk_1)(kl_1-k_1l),\label{omega}
\end{eqnarray}
under the parameter constraint
\begin{eqnarray}
(kl_1-k_1l)^2\sigma^2=3k_1^2(k_1^4c_0-kk_1^3c_1+k^2k_1^2c_2
-k_1k^3c_3+4k^4),\label{kl}
\end{eqnarray}
where
\begin{eqnarray}
&&c_0=4b^4+(a+bc)\left(4b^2\frac{a-bc}{c^2-1}
+\frac{a^2-b^2}{a-bc}\right),\nonumber\\
&&c_1=8b\frac{a^2+abc-2b^2}{1-c^2}-6ab\frac{c^2-1}{a-bc}
-4b-4ac,\nonumber\\
&&c_2=4b^2+4+(c^2-1)\frac{5a+bc}{a-bc}
+4\frac{a^2+4abc-5b^2}{c^2-1}\nonumber\\
&&c_3=-16b-2c\frac{c^2-1}{a+bc}-8c\frac{a-bc}{c^2-1}.
\end{eqnarray}
Especially, if we take $b=c=0$, then the solution \eqref{SW} becomes
\begin{eqnarray}
&&W=\frac{m}2\mbox{\rm sn}(X,\ m),\ p=\frac12\ln\left[\mbox{\rm dn}(X,\ m)-m\mbox{\rm cn}(X,\ m)\right],\label{sn}\\
&&\omega_1=\frac{k_1}k\omega-\frac{k_1^5m^2}{4k^2}-(m^2+1)k_1^3
-\frac{\sigma^2}{k^2k_1}(5kl_1+lk_1)(kl_1-lk_1)+12k^2k_1,\\
&&(kl_1-lk_1)^2\sigma^2=\frac34m^2k_1^6-3(m^2+1)k^2k_1^4+12k_1^2k^4.
\end{eqnarray}

Unfortunately, though \eqref{f} with \eqref{pW} and \eqref{SW} (and then \eqref{sn}) is a solution of the usual KP equation \eqref{KP}, it can not be used to find nontrivial solution of the ABKP equation \eqref{ABKP}.

It is interesting that some special solution of \eqref{Wx} can also be expressed by means of the Jacobi elliptic functions in some alternative ways, say,
\begin{eqnarray}
&&W=\frac{a\ \mbox{\rm sn}^2(X,\ m)+b}{1-c\ \mbox{\rm sn}^2(X,\ m)},\ a = -bc \pm \sqrt{c(c-1)(c-m^2)}, \label{SW1}\\
&&p=\left(b+\frac{a}c\right)E_{\pi}(\mbox{\rm sn}(X,\ m),c,m)-\frac{a}cX,\label{SP1}\\
 &&E_{\pi}(z,c,m)\equiv \int_0^z\frac{\mbox{\rm dt}}{(1-ct^2)\sqrt{(1-t^2)(1-m^2t^2)}} \label{pi}
\end{eqnarray}
with
\begin{eqnarray}
\omega_1=\frac{k_1}k\omega-\frac{k_1^5}{k^2}c_0+k_1^3c_2-2kk_1^2c_3
+12k_1k^2-\frac{\sigma^2}{3k_1k^2}(5kl_1+lk_1)(kl_1-k_1l),
\label{omega1}
\end{eqnarray}
under the parameter constraint
\begin{eqnarray}
(kl_1-k_1l)^2\sigma^2=3k_1^2(k_1^4c_0-kk_1^3c_1+k^2k_1^2c_2
-k_1k^3c_3+4k^4),\label{kl1}
\end{eqnarray}
where
\begin{eqnarray*}
&&c_3=\frac{4(2a^2+c^3-cm^2-2b^2c^2)}{c(a+bc)},\\
&&c_2=\left(\frac{a}c-2b\right)c_3+4c-12b^2
+\frac{8ab}c-\frac{4a^2}{c^2},\\
&&c_1=\left(\frac{ab}c-b^2-\frac{a^2}{c^2}\right)c_3
+\left(\frac{a}c-b\right)c_2
+\frac4{c^3}(a-bc)(a^2+b^2c^2),\\
&&c_0=\frac{a^3}{c^3}c_3-\frac{a^2}{c^2}c_2
+\frac{a}cc_1-\frac{4a^4}{c^4}.
\end{eqnarray*}

From solution \eqref{f} with \eqref{p} of the usual KP equation \eqref{KP}, we know that \eqref{f} is not a solution of the ABKP equation \eqref{ABKP} except that $f$ and $p$ is antisymmetric with respect to the operator $\hat{P}_s^y\hat{P}_s^x\hat{T}_d$. Thus the $\hat{P}_s^y\hat{P}_s^x\hat{T}_d$ invariant interaction solutions between soliton and $p$-wave possess the form
\begin{eqnarray}
A&=&-f_{x}^2\tanh^2(f)+f_{xx}\tanh(f)
-\frac{1}{12}\frac{f_{y}^2\sigma^2}{f_{x}^2}
+\frac{2}{3}f_{x}^2\nonumber\\
&&-\frac{1}{3}\frac{f_{xxx}}{f_{x}}-\frac{1}{12}
\frac{f_{t}}{f_{x}}+\frac{1}{4}\frac{f_{xx}^2}{f_{x}^2},\ f=k\left(x-\frac{x_0}2\right)+l\left(y-\frac{y_0}2\right)+\omega \left(t-\frac{t_0}2\right)+p, \label{f1}
\end{eqnarray}
with $p$ being a solution of the $p$-wave equation \eqref{p} under the antisymmetric condition
\begin{eqnarray}
\hat{P}_s^y\hat{P}_s^x\hat{T}_dp=-p. \label{p1}
\end{eqnarray}
For the travelling $p$-wave solution \eqref{SP1}, if we fix the constant $X_0=-\frac12(k_1x_0+l_1y_0+\omega_1t_0)$, then it will satisfy the antisymmetric condition \eqref{p1}, i.e.,
\begin{eqnarray}
p=\left(b+\frac{a}c\right)E_{\pi}(\mbox{\rm sn}(\xi,m), c, m)-\frac{a}c\xi,\ \xi=k_1\left(x-\frac{x_0}2\right)
+l_1\left(y-\frac{y_0}2\right)
+\omega_1\left(t-\frac{t_0}2\right),\label{sp1}
\end{eqnarray}
where ten constants $a,\ b,\ c,\ m,\ k,\ k_1,\ l,\ l_1,\ \omega$ and $\omega_1$ satisfy only three conditions \eqref{SW1}, \eqref{omega1} and \eqref{kl1}.

It is worth to mention that all the $\hat{P}_s^y\hat{P}_s^x\hat{T}_d$ invariant solutions obtained in this section are solutions not only for the ABKP equation \eqref{ABKP} but also for all ABKP equations
\begin{equation}
F(A,\ B)=0,\qquad F(u,\ u)=KP, \label{GABKP}
\end{equation}
where $KP$ is defined as \eqref{KP}.

For more concretely, the multiple soliton solutions \eqref{A13}, the Painlev\'e reductions \eqref{Aw} and the soliton-$p$-wave interaction solution \eqref{f1} with \eqref{p} and \eqref{p1} (and then the soliton-cnoidal wave interaction solution \eqref{f1} with \eqref{sp1}) are solutions for all ABKP equations \eqref{GABKP}, especially for
\begin{equation}
\begin{split}
&\left[A_t+\frac{f_1(A,B)}{f_1(B,A)}A_{xxx}
+\frac{3f_2(A,B)}{2f_2(A,B)}\left(A+\frac{f_3(A,B)}
{f_3(B,A)}B\right)
\left(B_x+3\frac{f_4(A,B)}{f_4(B,A)}A_x\right)\right]_x\\
&\quad
+\sigma^2\frac{f_5(A,B)}{f_5(B,A)}A_{yy}+{f_6(A,B)}-{f_6(B,A)}=0,
\label{ABKP1}\\
&B=\hat{P_s^x}\hat{P_s^y}\hat{T_d}A=A(-x+x_0,\ -y+y_0,\ -t+t_0)
\end{split}
\end{equation}
with arbitrary functionals $f_i(A,B),\ i=1,2,\ldots,6$.

It should be emphasized that the existence of multiple soliton solutions of \eqref{GABKP} and \eqref{ABKP1} does not imply the integrability of \eqref{GABKP} and \eqref{ABKP1}. In other words, the existence of multi-soliton solutions of a partial differential system is not a sufficient condition of the integrability of the related model.

 The $\hat{P}_s^y\hat{P}_s^x\hat{T}_d$ invariant solutions mean the event happened at $\{x,t\}$ will happen also at $\{x',t'\}\equiv\{-x+x_0,-t+t_0\}$. Then new problems arise up: Are there and how to obtain $\hat{P}_s^y\hat{P}_s^x\hat{T}_d$ symmetry breaking solutions of the ABKP system \eqref{ABKP}?

\section{$\hat{P}_s^y\hat{P}_s^x\hat{T}_d$ symmetry breaking solutions of the ABKP system \eqref{ABKP}}
In this section. we aim to search for the $\hat{P}_s^y\hat{P}_s^x\hat{T}_d$ symmetry breaking solutions of the ABKP system \eqref{ABKP}.
Motivated by the $\hat{P}_s^y\hat{P}_s^x\hat{T}_d$ symmetry invariant solutions of the nonlocal ABKP system \eqref{ABKP} being obtained from those of the usual local KP equation \eqref{KP}, we now construct the $\hat{P}_s^y\hat{P}_s^x\hat{T}_d$ symmetry breaking solutions of the nonlocal ABKP system \eqref{ABKP} from a coupled local KP system
\begin{equation}
\left[u_t+u_{xxx}+\frac{3}{2}(u+v)(v_x+3u_x)\right]_x+\sigma^2u_{yy}=0, \label{ckpu}
\end{equation}
\begin{equation}
\left[v_t+v_{xxx}+\frac{3}{2}(u+v)(u_x+3v_x)\right]_x+\sigma^2v_{yy}=0.
\label{ckpv}
\end{equation}
It can be directly found that the coupled KP system \eqref{ckpu}-\eqref{ckpv} possesses a special reduction of
\begin{equation}
v=B,\ u=A,\ B=A(-x+x_0, -y+y_0,\ -t+t_0),\label{uvAB}
\end{equation}
 which makes the coupled KP system \eqref{ckpu}-\eqref{ckpv} being reduced to the nonlocal ABKP equation \eqref{ABKP}.

Thus, to look for the exact solutions of the ABKP equation \eqref{ABKP} is equivalent to find out the exact solutions of the coupled KP equation system \eqref{ckpu}-\eqref{ckpv} with a special reduction condition \eqref{uvAB}.

In this section, we will show details on how to obtain the $\hat{P}_s^y\hat{P}_s^x\hat{T}_d$ symmetric breaking solutions of the ABKP system \eqref{ABKP} with the help of the coupled KP system \eqref{ckpu}-\eqref{ckpv}.

\subsection{Symmetry breaking multiple soliton solutions of the ABKP system \eqref{ABKP}}
As we have known, one of the notable features of the integrable system is possessing of multiple soliton solutions though it is not a sufficient condition. Many reliable methods are used in  literatures to examine the soliton solutions of integrable nonlinear evolution equations. The Hirota bilinear method\cite{ref19}, the B\"{a}cklund transformation method, the inverse scatting method\cite{ref20}, the Painlev\'{e} analysis, and others are effectively used to determine soliton solutions for completely integrable equations. Among all methods, the tanh and elliptic function expansion approach \cite{ref21,ref22,ref23,ref24} is one of the simplest and effective methods to search for one solitary wave and one periodic travelling wave for integrable and nonintegrable systems.

Because the $\hat{P}_s^y\hat{P}_s^x\hat{T}_d$ symmetry invariant multiple solitons have been obtained from the usual KP equation, it is reasonable to assume that the multiple soliton solutions of the coupled KP system \eqref{ckpu}-\eqref{ckpv} possess the form
\begin{equation}
u=\left(\ln \psi\right)_{xx}+a q_x,\ v=\left(\ln \psi\right)_{xx}-a q_x,\label{uvq}
\end{equation}
with a symmetry invariant part ($q$ independent part) and symmetry breaking part ($q$ dependent part),
where $\psi$ is a solution of the usual bilinear KP equation
\begin{eqnarray}
&&(D_xD_t+D_x^4+\sigma^2 D_y^2+(\lambda_1+\lambda_2)x+\lambda)\psi\cdot\psi=0,\label{BLKP}\\
&&D_xD_yD_t \psi\cdot \psi \equiv \left.(\partial_x-\partial_{x'})
(\partial_y-\partial_{y'})
(\partial_t-\partial_{t'})\psi(x,y,t)\psi(x',y',t')
\right|_{x'=x,y'=y,t'=t} \nonumber
\end{eqnarray}
and $\lambda,\ \lambda_1$ and $\lambda_2$ are arbitrary functions of $\{y,\ t\}$.
It is straightforward to check that substituting \eqref{uvq} into the coupled KP system \eqref{ckpu}-\eqref{ckpv} yields the equation system \eqref{BLKP} and
\begin{equation}
q_{xt}+\sigma^2q_{yy}+6(\ln \psi)_{xx}q_{xx}+q_{xxxx}=\frac12(\lambda_2-\lambda_1).\label{eqq}
\end{equation}
If
$$\lambda_2=-\lambda_1=\frac12\lambda,$$
\eqref{eqq} possesses a special solution
\begin{equation}
q=\ln \ \psi.
\end{equation}
Furthermore, if we take
\begin{equation}
\lambda_1=\lambda_2=\lambda=0,\label{ll}
\end{equation}
we can obtain a special multiple soliton solution of the coupled
KP equation system \eqref{ckpu}-\eqref{ckpv} in the form
\begin{eqnarray}
&& u=(\partial_x^2+a\partial_x)\ln F,\label{up}\\
&& v=(\partial_x^2-a\partial_x)\ln F,\label{vp}
\end{eqnarray}
where $F$ is given by \eqref{F}. Now, using the symmetry reduction condition \eqref{uvAB}, we get the
$\hat{P}_s^y\hat{P}_s^x\hat{T}_d$ symmetry breaking multiple soliton solution for the ABKP system \eqref{ABKP}
\begin{eqnarray}
&& A=(\partial_x^2+a\partial_x)\ln \sum_{\nu=-1,1}
K_\nu\cosh\left(\frac12\sum_{j=1}^N\nu_j\eta_j\right),
\label{Ap}\\
&&K_\nu=\prod_{i>j}\sqrt{3k_i^2k_j^2(k_i-\nu_i\nu_jk_j)^2
-\sigma^2(l_ik_j-k_il_j)^2},\ \nonumber\\
&&\eta_j=k_j\left(x-\frac{1}{2}x_0\right)
+l_j\left(y-\frac{1}{2}y_0\right)-k_j^{-1}\left(k_j^4
+\sigma^2l_j^2\right)\left(t-\frac{1}{2}t_0\right),\nonumber
\end{eqnarray}
where the summation of $\nu=\{\nu_1,\ \nu_2,\ \ldots,\ \nu_N\}$ should be done for all non-dual permutations of $\nu_i=1,\ -1,\ i=1,\ 2,\ \ldots,\ N$.

It is clear that when the symmetry braking parameter $a$ is fixed as zero, the symmetry breaking multiple soliton solution \eqref{Ap} is reduced back the symmetry invariant solution \eqref{A13} given in the last section.

To get more general symmetry breaking multiple soliton solutions of the coupled KP and then the nonlocal ABKP systems, we have to study the general solutions of \eqref{BLKP} and \eqref{eqq}.
Here, we just give a discussion for the single soliton case with the condition \eqref{ll}.

For the single soliton solution of \eqref{BLKP} with \eqref{ll}, we have
\begin{equation}
\psi=\cosh \frac{\xi}2,\ \xi=kx+ly-\frac{k^4
+\sigma^2l^2}{k} t+\eta_0 \label{ones}
\end{equation}
and the $q$ equation \eqref{eqq} becomes
\begin{equation}
q_{xt}+\sigma^2q_{yy}+\frac32 \mbox{\rm sech}^2 \frac{\xi}2 \ q_{xx}+q_{xxxx}=0\label{eq1}
\end{equation}
which can be solved via variable separation approach
\begin{equation}
q=\sum_{i=1}^M \Xi_i(\xi)Y_i(y,\ t),\ \label{aq}
\end{equation}
where $\Xi_i$ and $Y_i$ are solutions of
\begin{equation}
\Xi_i''''+\frac12\left[1-3\tanh^2\left(\frac{\xi}2\right) \right]\Xi_i''+\beta_i \Xi'_i+\alpha_i\Xi_i=0,  \label{Xi}
\end{equation}
and
\begin{equation}
Y_{i,yy}=-\alpha_i^2\sigma^2 k^4Y_i,\ Y_{i,t} = \beta_i k^3 Y_i-2l\sigma^2k^{-1}Y_{i,y}  \label{Yi}
\end{equation}
with arbitrary variable separated parameters $\alpha_i$ and $\beta_i$.

The general solution of \eqref{Yi} can be written as
\begin{equation}
Y_{i}=\left\{\begin{array}{ll}
(c_ik y-2c_il\sigma^2t+d_i)e^{\beta_ik^3t}, & \alpha_i=0 \\
c_i\sin[\alpha_ik(k\sigma y-2l\sigma^{-1}t+d_i)]e^{\beta_ik^3t}, & \alpha_i\neq0
\end{array}\right.  \label{rYi}
\end{equation}
with arbitrary constants $c_i$ and $d_i$.

Though the $\Xi$ equation \eqref{Xi} is only a variable coefficient linear ordinary equation, it is still very difficult to get general solution for $\alpha_i\neq0$. Here, we write down only a special case for $\alpha_i=\beta_i=0$,
\begin{equation}
\Xi_i=c_{1}+c_2\xi+c_3\ln\left(\cosh \frac{\xi}2\right)
+c_4\int \left[\cosh(\xi)+3\xi\tanh\left(\frac{\xi}2\right)\right]\mbox{\rm d}\xi
     \label{rXi}
\end{equation}
Combining the solutions \eqref{uvq},\ \eqref{aq}, \eqref{rYi}, \eqref{rXi} and the reduction condition \eqref{uvAB}, we get a single symmetry breaking soliton solution
\begin{eqnarray}
A&=&-k^2\mbox{\rm sech}^2\ \frac{\xi}2 +\eta(b_0 +b_2\cosh\ \xi)
+(b_1+b_2\xi\eta) \tanh \ \frac{\xi}2,\ \eta =ky' -2l\sigma^2 t',\label{SB}\\
\xi&=&kx'+ly'
-\frac{k^4+\sigma^2l^2}kt',\
\ x'\equiv x-\frac{x_0}2,\ y'\equiv y-\frac{y_0}2,\ t'\equiv t-\frac{t_0}2 \label{xy'}
\end{eqnarray}
with arbitrary constants $a_0,\ a_1,\ a_2,\ b_0,\ b_1$ and $b_2$.
The solution \eqref{SB} is $\hat{P}_s^y\hat{P}_s^x\hat{T}_d$ symmetry breaking for $b_0^2+b_1^2+b_2^2\neq0$ and symmetry invariant for $b_0=b_1=b_2=0$. The symmetry breaking solution is analytic only for $b_0=b_2=0$.

\subsection{$\hat{P}_s^y\hat{P}_s^x\hat{T}_d$ symmetry breaking cnoidal wave solutions}
In this subsection, we try to explore the $\hat{P}_s^y\hat{P}_s^x\hat{T}_d$ symmetry breaking cnoidal wave solutions for the ABKP system \eqref{ABKP}. As stated before, we first start from the coupled KP system \eqref{ckpu}-\eqref{ckpv} to construct the conidal wave solutions, and then apply the constraint reduction $v=\hat{P}_s^y\hat{P}_s^x\hat{T}_du$ to these solutions of coupled KP system \eqref{ckpu}-\eqref{ckpv} to find the solutions of the ABKP system \eqref{ABKP}.

In fact, because any function can be separated to the summation of the symmetry and antisymmetry parts. Thus we can always write the solutions of the coupled KP system \eqref{ckpu}-\eqref{ckpv} in this way
\begin{equation}
u=w+z_x,\ v=w-z_x,\label{wz}
\end{equation}
where $w$ and $z$ present $\hat{P}_s^y\hat{P}_s^x\hat{T}_d$ symmetric and antisymmetric part respectively. As mentioned in section III, for the ABKP systems,  $\hat{P}_s^y\hat{P}_s^x\hat{T}_d$ symmetric solutions should be special solutions of the usual KP equation. Substituting \eqref{wz} into the coupled KP system \eqref{ckpu}-\eqref{ckpv} we find that $w$ is a solution of the usual KP equation \eqref{KP} while $z$ should be solution of the following $z$ equation
\begin{equation}
z_{xt}+z_{xxxx}+\sigma^2 z_{yy}+6wz_{xx}=0. \label{zxt}
\end{equation}
For the cnoidal wave solution of the usual KP equation, we take
\begin{equation}
w=-\frac14k^2 m^2 \mbox{\rm sn}^2 \left(\frac12\xi\right), m)^2+\frac{k^2}8(m^2+1),\ \xi=kx+ly-\frac1{2k}[k^4(1+m^2)+2l^2\sigma^2]. \label{cw}
\end{equation}
Substituting \eqref{cw} into \eqref{zxt} and using variable separation approach, one can find a special solution
\begin{eqnarray}
z_\xi&=&[a_1+b_1(ky-2l\sigma t)]\mbox{\rm sn}\left(\frac{\xi}2,\ m\right)+[a_2(ky-2l\sigma^2t)+b_2] \nonumber\\
&&+[a_3+b_3(ky-2l\sigma^2t)]\int H\left(1-m^2,\frac{m^2}4-\frac54,-1,\frac32,\frac12,\frac12,\mbox{\rm dn}^2 \frac{\xi}2 \right)\mbox{\rm d}\xi\nonumber\\
&&+[a_4+b_4(ky-2l\sigma^2t)]\int H\left(1-m^2,-\frac34,-\frac12,2,\frac32,\frac12,\mbox{\rm dn}^2 \frac{\xi}2 \right)\mbox{\rm dn}\ \frac{\xi}2\mbox{\rm d}\xi,
 \label{cz}
\end{eqnarray}
where $a_i$ and $b_i,\ 1,\ 2,\ 3,\ 4$ are arbitrary constants and  the function $H=H(a,q,\alpha,\beta,\gamma,\delta,x)$ is defined as the known Heun general function which is a solution of the Heun equation
\begin{eqnarray}
&&H_{xx}=\left[\frac{(\alpha+\beta+1)}{a-x}
-\frac{a\gamma}{x(a-x)}-\frac{(a-1)\delta}{(x-1)(a-x)}\right]H_x
+\frac{(\alpha\beta x-q)H}{x(x-1)(x-a)},\label{H}\\
&&H(0)=1,\ H_x(0)=\frac{q}{\gamma a}. \nonumber
\end{eqnarray}
Finally, using the symmetry reduction condition, $v=\hat{P}_s^y\hat{P}_s^x\hat{T}_du$,
we get a special symmetry breaking cnoidal solution
\begin{eqnarray}
A&=&\frac{k^2}8(m^2+1)-\frac14k^2 m^2 \mbox{\rm sn}^2 \left(\frac12\xi\right), m)^2+a_1\mbox{\rm sn}\left(\frac{\xi}2,\ m\right)+a_2(ky'-2l\sigma^2t') \nonumber\\
&&+a_3\int^\xi H\left(1-m^2,\frac{m^2}4-\frac54,-1,\frac32,\frac12,\frac12,\mbox{\rm dn}^2 \frac{x}2 \right)\ \mbox{\rm dx}\nonumber\\
&&+a_4\int^\xi H\left(1-m^2,-\frac34,-\frac12,2,\frac32,\frac12,\mbox{\rm dn}^2 \frac{\xi}2 \right)\mbox{\rm dn}\ \frac{x}2\ \mbox{\rm dx}
 \label{cA}
\end{eqnarray}
for the ABKP equation \eqref{ABKP}, where $\{\xi,\ y',\ t'\}$ are defined by \eqref{xy'}. When $a_1=a_2=a_3=a_4$ the $\hat{P}_s^y\hat{P}_s^x\hat{T}_d$ symmetry breaking solution \eqref{cA} reduces back to the symmetry invariant periodic wave solution.

\section{Summary and discussion}
In summary, a special ABKP system is directly obtained from KP equation to describe two-place physical events by using consistent correlated bang. The ABKP system possesses $\hat{P}_s^y\hat{P}_s^x\hat{T}_d$ symmetry which means the ABKP system is invariant under the transformation $\{x\rightarrow-x+x_0,\ y\rightarrow -y+y_0,\ t\rightarrow-t+t_0\}$. The ABKP system is nonlocal and can be used to describe some special two-place physical problems.

With the help of the usual local KP equation and a local coupled KP system, we obtained some types of exact $\hat{P}_s^y\hat{P}_s^x\hat{T}_d$ invariant and $\hat{P}_s^y\hat{P}_s^x\hat{T}_d$ symmetry breaking solutions for the ABKP systems with different methods. The $\hat{P}_s^y\hat{P}_s^x\hat{T}_d$ invariant solutions of the ABKP systems \eqref{ABKP}, \eqref{ABKP1} and \eqref{GABKP}, such as multiple soliton solutions, soliton $p$-wave interaction solutions and symmetry reduction solutions (Painlev\'e IV, III and II reductions) are obtained from those of the usual local KP equation \eqref{KP} by the $\hat{P}_s^y\hat{P}_s^x\hat{T}_d$ invariant principle. The $\hat{P}_s^y\hat{P}_s^x\hat{T}_d$ symmetric breaking solutions, such as the multiple soliton solutions and periodic waves are obtained from a coupled KP system which show rich structures of the ABKP system.

Though the results of the nonlocal ABKP system \eqref{ABKP} are obtained via the usual local KP equation \eqref{KP} and the local coupled KP system \eqref{ckpu}-\eqref{ckpv}, the solutions of the nonlocal ABKP \eqref{ABKP} possess much more abundant structure than those of the local KP equation \eqref{KP} even if for the single soliton and single travelling periodic wave.

In this paper, we investigate only some types of special solutions of a special nonlocal ABKP equation \eqref{ABKP}. There are various important problems on the nonlocal multi-place KP systems should be deeply studied. In fact, there are some different types of non-localities such as those pointed out in \eqref{G} and there are also many other types of nonlocal KP equations such as those listed in \cite{ref15} and  equations \eqref{GABKP}, \eqref{ABKP1} and \eqref{Bang}. Many four place nonlocal KP systems have also been given by one of us (Lou), for instance, one set of integrable two-place and four place KP systems can be written as
\begin{eqnarray}
&&q_{xt}+\left(q_{xx}+\frac32\frac{\parallel u\parallel^2}{a-1}+6uq-3\beta vw \right)_{xx}+3\sigma^2q_{yy}=0,\\
&&u\equiv (1+\hat{f}+\hat{g}+\hat{f}\hat{g})q,\ \parallel u\parallel^2\equiv (1+\hat{f}+\hat{g}+\hat{f}\hat{g})q^2,\nonumber\\
&&
v\equiv (a+\hat{f})(1+\hat{g})q,\ w\equiv (1+\hat{f})(1-\hat{g})q \nonumber
\end{eqnarray}
with arbitrary constants $a$ and $\beta$ and $ \hat{f},\hat{G}\in {\mbox{G}}$ while the symmetry group ${\mbox{G}}$ is defined in \eqref{G}.

Because there exist various two-place and multi-place correlated physical events in almost all natural scientific fields, the multi-place physical problems and multi-place mathematical models should be attracted more attentions.

\section*{Acknowledgements}
The work was supported by NNSFC (No. 11435005). And the authors were sponsored by K. C. Wong
Magna Fund in Ningbo University.

\section*{}
\leftline{\bf Conflict of Interest: \rm The authors declare that they have no conflict of interest.}
\bf Funding: \rm This study was funded by the National Natural Science Foundation of China (grant number. 11435005).


\begin{thebibliography}{99}
\bibitem{ref1}M. J. Ablowitzl and Z. H. Musslimani, Phys. Rev. Lett. 110(2013)064105.

\bibitem{ref2}C. Q. Song, D. M. Xiao and Z. N. Zhu, Commun. Nonl. Sci. Numer. Simul. 47(2017)1.

\bibitem{ref3}S. Modak, A. Singh and P. Panigrahi, Eur. Phys. J. B 89(2016)149.

\bibitem{ref4}M. J. Ablowitzl and Z. H. Musslimani, Nonlinearity 29(2016)915.

\bibitem{ref5}J. L. Ji and Z. N. Zhu, J. Math. Anal. Appl. 453(2017)973.
%Soliton solutions of an integrable nonlocal modified Korteweg-de Vries equation through inverse scattering transform,arXiv:1603.03994,2016;


\bibitem{ref6}M. J. Ablowitzl and Z. H. Musslimani, Phys. Rev. E 90(2014)032912.

\bibitem{ref7}M. Dimakos and A. S. Fokas, J. Math. Phys. 54(2014)081504.

\bibitem{ref8}A. S. Fokas, Phys. Rev. Lett. 96(2006)190201.

\bibitem{ref9}A. S. Fokas, Nonlinearity, 29(2016)319.

\bibitem{ref10}C. M. Bender, Rep. Prog. Phys. 70(2007)947.

\bibitem{ref11}K. G. Markis, R. Ei-Ganainy, D. N. Christodoulidues and Z. H. Musslimani, Phys. Rev. Lett. 100(2008)103904.

\bibitem{ref12}C. M. Bender and S. P. Klevansky, Phys. Rev. Lett. 105(2015)031601.

\bibitem{ref13}Z. Lin, J. Schindler, F. M. Ellis and T. Kottos, Phys. Rev. A 85(2012)050101.

\bibitem{ref14}T. A. Gadzhimuradov and A. M. Agalarov, Phys. Rev. A, 93(2016)062124.

\bibitem{ref15}S. Y. Lou, Alice-Bob systems, $P_s$-$T_d$-$C$ principles and multi-soliton solutions, arXiv: 1603.03975v2[nlin.SI], 2016; J. Math. Phys. 59 (2018) 083507.
\bibitem{BYang}
B. Yang and Y. Chen, Nonl. Dyn. 94(2018)389;  Chaos, 28(2018)053104;
\bibitem{ref25}M. Jia and S. Y. Lou,
%Exact Ps Td, invariant and Ps Td, symmetric breaking solutions, symmetry reductions and B\"acklund transformations for an AB¨CKdV system[J].
Phys. Lett. A,, 382(2018)1157.
%-1166.
\bibitem{ref16}S. Y. Lou, Chin. Phys. Lett. 34(2017)060201 ;
S. Y. Lou and F. Huang, Sci. Rep.7(2017)869.


\bibitem{kp1} B. B. Kadomtsev and V. I. Petviashvili, Sov. Phys. Dokl.  15(1970) 539.
\bibitem{kp2} M. J. Ablowitz and H. Segur, J. Fluid Mech. 92(1979)691.
\bibitem{kp3} D. E. Pelinovsky, Yu. A. Stepanyants, and Yu. A. Kivshar, Phys. Rev. E 51(1991)5016.

\bibitem{ref17}R. Hirota, Phys. Rev. Lett. 27£¨1971£©1192.

\bibitem{ref18}S. Lie, Vorlesungen \"{u}ber Differentialgleichungen mit Bekannten Infinitesimalen Transformationen, Teuber, Leipzig, 1981 (1967 reprinted by Chelsea, New York).

\bibitem{ref28}P. J. Olver, application of Lie Groups to Differential Equation, 2nd ed. Graduate Texts in Mathematics, Springer-Verlag, New York, 1993.

\bibitem{ref29}G. W. Blumam and S. Kumei, Symmetries and Differential Equations, Appl. Math. Sci., Springer-Verlag, Berlin, 1989.

\bibitem{ref27}
S. Y. Lou, J. Phys. A: Math. Gen. 23 (1990) L649.

\bibitem{ref26}B. Ren,
%CTE method and interaction solutions for the Kadomtsev-Petviashvili equation[J].
J. Kor. Phys. Soc., 70(2017)333.
%-338.

\bibitem{ref30}S. Y. Lou, Stud. Appl. Math. 134(2015)372.

\bibitem{Phi3} S. Y. Lou,  G. X. Huang and G. J. Ni,
%Transforming some special solutions of  ¦Ë¦Õ4 model to that of ¦µ6 and ¦µ4+¦µ3 models,
Phys. Lett. A 146 (1990) 45.
%-49.


\bibitem{ref19}R. Hirota, J. Math. Phys. 14(1973)805.%-810.

\bibitem{ref20}M. J. Ablowitzl and P. A. Clarkson, Solitons, Nonlinear Evolution Equations and Inverse Scatting Transform, University Press, Cambridge, 1991.

\bibitem{ref21}H. B. Lan, K. L. Wang, J. Phys. A, Math. Gen. 23(1990)4097.

\bibitem{ref22}E. G. Fan, Phys. Lett. A 277(2000) 212-218.

\bibitem{ref23}A. Khare, and A. Saxena, J. Math. Phys. 55(2014) 032701.

\bibitem{ref24}S. Y. Lou, and G. J. Ni, J. Math. Phys. 30(1989)1614.


\end{thebibliography}
\end{document}